# The Search and Study of Low-Mass Scalar $\sigma_0$-mesons at the Impulse of Neutron Beam $P_n$ = 3.83 GeV/c

Yu.A.Troyan, A.V.Beljaev, A.Yu.Troyan, E.B.Plekhanov, A.P.Jerusalimov, S.G.Arakelian

## Introduction

This work is devoted to search and study scalar $0^+ \left[0^{++}\right]$ $\sigma_0$ – mesons in the system of $\pi^+\pi^-$ from the reaction $np \rightarrow np\pi^+\pi^-$ at the impulse of the quasimonochromatic neutrons $P_n = 3.83\, GeV/c$ from the data obtained in an exposure of the 1m $H_2$ bubble chamber of LHE (JINR).

To emphasize importance of the given research, we shall quote the statement of Dr. M.R.Pennington: "So why are the light scalars interesting? This is because they are fundamental. They constitute the Higgs sector of the strong interaction. It is these scalar fields, which have a non-zero vacuum expectation value that breaks chiral symmetry and ensures pions are very light, while giving mass to all other light flavored hadrons" [1].

Scalar meson with weight, smaller than $1\, GeV/c^2$ can belong to family of Higgs bosons (so-called Higgs bosons of strong interactions). Search and finding-out of their properties is, now, the central problem of physics of particles.

In our works [2] there are reviews of theoretical approaches to this theme and of experimental results which are accessible in present time.

As following from theoretical works, understanding of the structure and properties of $\sigma_0$–mesons is not present yet. Therefore the careful research of these questions is extremely important.

## The reaction $np \rightarrow np\pi^+\pi^-$ at $P_n = 3.83\, GeV/c$

VBLHE JINR is an unique laboratory in which there are data of neutron-proton interactions with various energies of neutrons. These data have been obtained from experiments with one-meter hydrogen bubble chamber (HBC), irradiated by quasi-monochromatic neutron beams ($\Delta P_n / P_n \approx 2.5\%$, $\Delta\Omega_{beam} = 10^{-7}\, sterad$.) [3]. There were organized nine experiments with the impulses of the neutron beam 1.25, 1.32, 1.43, 1.73, 2.25, 3.83, 4.42, 5.20 $GeV/c$. HBC has been used as tracks detector and as liquid-hydrogen target. In total, near a half of million of events np - interactions are accumulated. Accuracy of definition of impulse of track in the HBC is (2÷3)%, an angular accuracy is ~0,5°.

The events from different reactions (without neutral particles and with one neutral particle) were separated using standard method $\chi^2$ with corresponding number of degree of freedom [4,5].

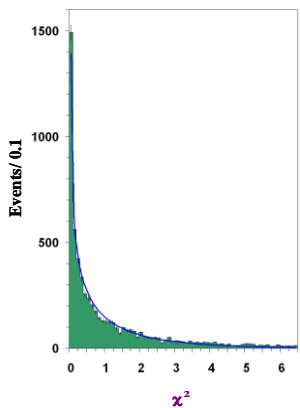
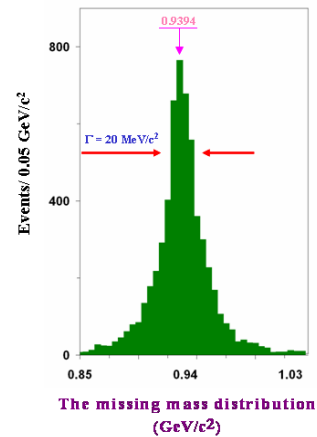

Fig.1 the $\chi^2$ distribution for the reaction $np \rightarrow np\pi^+\pi^-$ at $P_n = 3.83\, GeV/c$
the histogram – the experimental distribution;
the curve - the theoretical distribution of $\chi^2$ with one degree of freedom

Fig.2 the missing mass distribution for the reaction $np \rightarrow np\pi^+\pi^-$ at $P_n = 3.83\, GeV/c$

Reaction $np \rightarrow np\pi^+\pi^-$ at $P_n = 3.83\, GeV/c$ has been chosen by method $\chi^2$ with one degree of freedom ($\chi^2 \leq 6.5$). Fig.1 shows the $\chi^2$ distribution for this reaction. In the same figure is shown the theoretical distribution of $\chi^2$ with one degree of freedom. Concurrence of both distributions is visible in the limits of mistakes of measurements. Fig.2 shows the missing mass distribution for this reaction. Its central value complies with mass of the neutron. The relations of the flight of neutrons, protons, $\pi^+$ and $\pi^-$ into the corresponding angular intervals in c.m.s. of the reaction are calculated. They, in the limits of mistakes, correspond to predicting by isotopic invariance for this reaction. The distributions of momenta of the particles in c.m.s. of the reaction have been exploring too. They also satisfy predictions of isotopic invariance in the given reaction.

In total, there are 600 events from the reaction $np \to np\pi^+\pi^-$ at $P_n = 3.83\,GeV/c$.

## Spectrum of the effective masses of $\pi^+\pi^-$

Fig.3 shows the distribution of the effective masses of $\pi^+\pi^-$. The strong effect with mass $M_{\pi^+\pi^-} = 404 \pm 3\,MeV/c^2$ and width $\Gamma_{Res}^{\exp} = 14 \pm 3.8\,MeV/c^2$ exists. The background, in this instance, is defined by mixing of $\pi^+$ and $\pi^-$ - mesons from decays of $\Delta_{33}^{++}$ and $\Delta_{33}^{-}$ - isobars, which plentifully are born in the given reaction. The background curve, taken as Legendre polynomial of 9 power, is shown. The parameters of the description of the background part of the distribution (without peak) are $\chi^2 = 0.85 \pm 0.19$, $\sqrt{D} = 1.41 \pm 0.13$. The description of the distribution by superposition of polynomial background and resonance curve, taken in Breit-Wigner form, is also given on the picture.

In the same picture the background distribution which is calculated by means OPER - model [6] is presented. This background is close to polynomial. The OPER (Reggeized One Pion Exchange) model is used for the description of reactions with a plural birth of π – mesons in the πN and NN - interactions. Unlike other models of peripheral processes, such as an exchange of elementary or virtual π - meson (including models with absorption), OPER model considers exchange a Regge-paths. The model including the processes of birth of basic barionic $N^*$ and $\Delta^*$ - resonances and the processes of elastic $\pi\pi \to \pi\pi$ dispersion. In addition, for the description of the reaction, the processes of difraction birthes of resonances $N^*_{1440}$, $N^*_{1520}$, $N^*_{1680}$ were used.

Advantages of OPER - model are:
- wide range of described energies ((2÷200 GeV);
- a small amount of the fixed free parameters (3 in our case;
- the calculated values are automatically normalized on the reaction cross-section.

Number of standard deviation from background for the existing resonance is $S.D. = (N_{Res} - N_{back})/\sqrt{N_{back.}}$ (where: $N_{Res}$ - number of events in the resonance region; $N_{back}$ - number or events in the resonance region, under the background curve) and it is $S.D. = 4.2$.

Probability that the given peak is an fluctuation of the background, does not exceed 0.001

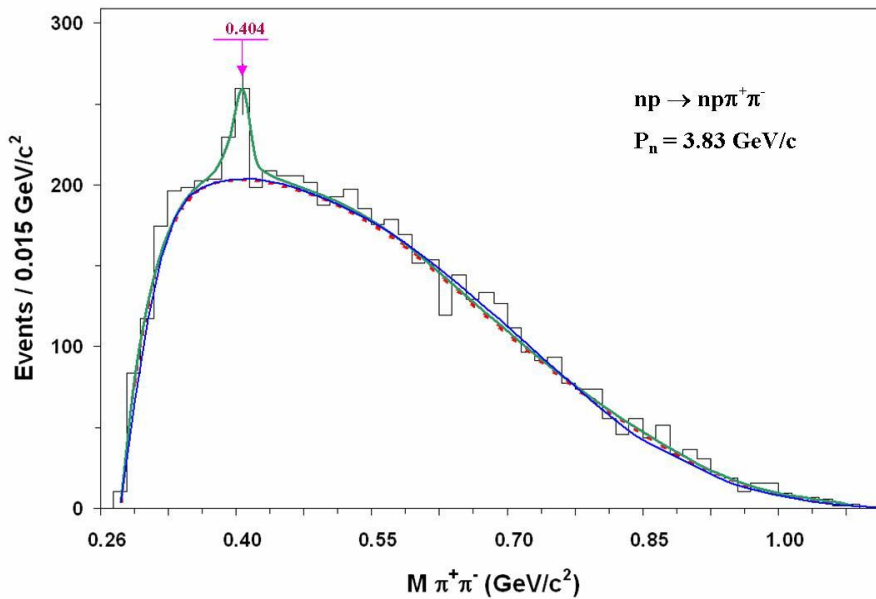

Fig.3 the distribution of the effective masses of of $\pi^+\pi^-$

red curve - the background curve, taken as Legendre polynomial green curve – the description of the distribution by superposition of polynomial background and Breit-Wigner resonance curve

blue curve - the background curve, taken by means OPER-model

## The estimation of the quantum numbers of the resonance

The quantum numbers of the existing resonance are estimating by next procedures:

The value of spin. For the spin estimation, firstly, the angular distribution of $\pi^+$ - mesons from resonance region, in the helicity system, has been constructed. By the same manner, secondly, has been constructed the distribution for the left and right side beside resonance region. Thirdly, the result of the subtraction of the second (background) from first (total) distribution - the distribution for the resonance - has been described by the set of Legendre polynomials of even powers, with maximum power being equal to $2k$. The spin $J$ of the resonance has been estimate as $J \geq k$. [7].

Several variants of distribution (more then 100 for existing resonance), with different borders of a resonance region and of a background region, has been constructed and tested. Considering, for each distribution, value of $\chi^2$, dispersion and probability of the description, gives the $2k$ as the maximal power in the most suitable set of Legendre polynomials.

Fig.3 shows the resulting distribution for the existing resonance. This distribution is isotropic. Consequently the value of the resonance spin, with high probability, $J_{Res} = 0$.

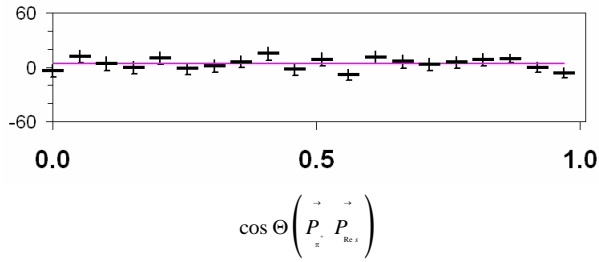

Fig.4   the distribution for the spin estimation for the existing resonancthe
direct red line – the most probable description (isotropic)

This means that the orbital momentum $l$ of the resonance is $l=0$, too.

P - parity. $P = (-1)(-1)(-1)^l = +1$ (with $l=0$).

C - parity. $C = (-1)^{l+s}$, where: $s = J_{\pi^+} + J_{\pi^-} = 0$, $l=0$. This way, $C = +1$.

G - parity. The G - parity for the system, which decays to π - mesons, is $G = (-1)^n$, where $n$ – an amount of the rotations in the charging space, necessities for return system in source condition. In the event of $\pi^+\pi^-$ - system $n=2$ and, therefore, $G = +1$.

Isotopic spin $I$. It is known that $G = C(-1)^I$, where $I$ - isotopic spin of the system. From $G = +1$ and $C = +1$, in our system, follows that $I=0$ or $I=2$.

For unequivocal choice of $I$, the system of resonances in $\pi^-\pi^-$-system from the reaction $np \to pp\pi^+\pi^-\pi^-$ at $P_n = 5.20\,GeV/c$ was considered [8]. There is the strong peak with $M_{\pi^-\pi^-} = 397\,MeV/c^2$ in the distribution of the effective masses of $\pi^-\pi^-$, that close to mass of investigated peak in the system of $\pi^+\pi^-$. However, estimation of the spin of resonance in system $\pi^-\pi^-$ gives the value $J \geq 6$. Thereby, this resonance has no relation to the peak, observed in the system of $\pi^+\pi^-$. From here, the value of isotopic spin for resonance $M_{\pi^+\pi^-} = 404\,MeV/c^2$ is $I=0$.

From the previous analysis, it is possible to draw a conclusion, that, with the big share of the probability, the observable resonance has quantum numbers of $\sigma_0$ – meson $0^+ \left[0^{++}\right]$.

## The results of investigation and the conclusion.

The table of results of investigation for meson with $M_{\pi^+\pi^-} = 404\,MeV/c^2$ from the reaction $np \to np\pi^+\pi^-$ at $P_n = 3.83\,GeV/c$ is presented.

In this table are:

$M_{Res}$ and $\Gamma_{Res}$ – experimental values of the mass and width of the resonance;

$\Gamma_{Res}^{true}$ – true width of the resonance, calculated with the account of mass resolution;

*S.D.* – number of standard deviation from background;

$\sigma_{\mu b}$ – cross-section of the resonance, calculated with the account of the reaction cross-section $\sigma_{np \to np\pi^+\pi^-} = (6.46 \pm 0.32) mb$ [5].

The mass resolution is calculated by formula

$\Gamma_{res}(M_{Res}) = 4.2 \cdot \left[\left(M_{Res} - \sum_{i=1}^{2} m_i\right)/0.1\right] + 2.8$ $(MeV/c^2)$, where $M_{Res}$ and $m_i$ (the masses of resonance and particles – components) are set in $GeV/c^2$, and equal, in the resonance region, is $\Gamma_{res} = 7.8\ MeV/c^2$.

The true width of the resonance is calculated by formula $\Gamma_{Res}^{true} = \sqrt{\left(\Gamma_{Res}^{exp}\right)^2 - \left(\Gamma_{res}\right)^2}$.

The table of results

| $M_{Res} \pm \Delta M_{Res}$ MeV/c² | $\Gamma_{Res}^{exp} \pm \Delta\Gamma_{Res}^{exp}$ MeV/c² | $\Gamma_{Res}^{true}$ MeV/c² | **S.D.** | $\sigma_{\mu b}$ |
|---|---|---|---|---|
| 404 ± 3 | 14 ± 3.8 | 10.4 | 4.2 | 86 ± 32 |

The σ₀ – mesons can be a powerful facility of the study of a hot and dense matter. As shown in paper [10], there are decreases of the weight and width of mesons in a hot and dense matter. Therefore search of such effects in nuclear interactions will help to observe a condition of a quark-gluon plasma. It is probably to make within the limits of projects NIKA/MPD and CBM.

Chamber np - data and pp - data from "HADES" allow to calculate a background from NN - interactions which are an overwhelming part of the effects appreciable in nuclear reactions

Authors thanks Dr. V.L. Lyuboshitz, Dr. A.I. Malakhov, Dr. M.V Tokarev, Dr. Yu.V. Zanevsky for the help in the and the useful discussions.

The investigation has been performed at the Veksler and Baldin Laboratory of High Energies, JINR (theme 1087).